\documentclass{article}
\usepackage[ruled,linesnumbered]{algorithm2e}
\usepackage{algorithmic}
\usepackage{lineno,hyperref}
\usepackage{graphicx}% Include figure files
\usepackage{dcolumn}% Align table columns on decimal point
\usepackage{bm}% bold math
\usepackage{color}% bold math
\usepackage{amsmath}
\usepackage[utf8]{inputenc}
\usepackage[T1]{fontenc}
\usepackage{graphicx}
\usepackage{hyperref}
\usepackage{epstopdf}
\usepackage{multirow, multicol,times,subfigure, setspace, rotating, url}
\usepackage{tabu}                     % 表格插入
\usepackage{multirow}                 % 一般用以设计表格，将所行合并
\usepackage{multicol}                 % 合并多列
\usepackage{multirow}                % 合并多行
\usepackage{float}                    % 图片浮动
\usepackage{makecell}                 % 三线表-竖线
\usepackage{booktabs}                 % 三线表-短细横线
\usepackage{amssymb}
\usepackage{pifont}
\usepackage{enumerate}
\usepackage{amsmath}
\usepackage[superscript, biblabel, nomove]
{cite}
\newcommand{\customcup}[2]{\bigcup\limits_{#1}^{#2}}

\begin{document}
% \title{Temporal network comparison based on spectral entropy}
\title{BCIM: Budget and capacity constrained influence maximization in multilayer networks}
\author{\normalsize
Su-Su Zhang{\small$^{\mbox{1}}$},
Chuang Liu{\small$^{\mbox{1}}$},
Huijuan Wang{\small$^{\mbox{3}}$},
Yang Chen{\small$^{\mbox{4,*}}$},
Xiu-Xiu Zhan{\small$^{\mbox{1,2,*}}$}
}

\maketitle

\vspace{-5mm}

% \noindent
$^1$ Research Center for Complexity Sciences, Hangzhou Normal University, Hangzhou 311121, PR China \\
$^2$ College of Media and International Culture, Zhejiang University, Hangzhou 310058, PR China\\
$^3$ Faculty of Electrical Engineering, Mathematics, and Computer Science, Delft University of Technology, Mekelweg 4, 2628 CD, Delft, The Netherlands\\
$^4$ Shanghai Key Lab of Intelligent Information Processing, School of Computer Science, Fudan University, Shanghai 200433, PR China\\
%* zhanxiuxiu@hznu.edu.cn

%\affil[+]{these authors contributed equally to this work}

%\keywords{Keyword1, Keyword2, Keyword3}

\begin{abstract}
Influence maximization (IM), which aims to select a set of seed nodes that can maximize their influence within a network, can be applied to various domains, such as viral marketing, disease control, and political campaigns. One of the extended IM problems, i.e., budgeted Influence Maximization (BIM), adds a budgeted constraint, which considers that different nodes may have different costs. However, the definition of the current BIM problem has limitations, as it only imposes a budget constraint. This may lead to solutions with a large number of low-cost nodes that are not applicable to real-world scenarios.
Furthermore, since users can transmit information on multiple social networking platforms, solving the BIM problem across different platforms could better optimize resource usage. To address these issues, we propose a Budget and Capacity Constrained Influence Maximization (BCIM) problem in multilayer networks and design a Multilayer Multi-Population Genetic Algorithm (MMGA) to solve it. In MMGA, a repair module addresses budget constraints, while a multi-population parallel module handles capacity (seed size) constraints. Additionally, the initialization module, which considers multilayer network properties, and the multi-population parallel module significantly enhance the algorithm's efficiency. Extensive experiments upon various synthetic and empirical multilayer networks indicate that MMGA achieves at least a 10\% improvement in spreading performance under budget and capacity constraints compared to baseline methods extended from various classical IM problems. The proposed BCIM problem opens up a new direction in influence maximization, and the proposed method offers an effective and efficient framework to the problem.

% new 

% However, the definition of the current BIM problem has limitations, that is, it only imposes a budget constraint, which may lead to solutions with a large number of low-cost nodes that are not applicable to real-world scenarios.

% However, traditional BIM problems overlook practical factors including capacity limitations and diverse interactions. Therefore, we introduce and resolve the budget and capacity constrained influence maximization problem (BCIM) in multilayer networks. 

% Previous explorations of maximizing influence have been conducted on single-layer networks without considering budget costs. Therefore, the problem of maximizing budget influence on multi-layer networks has gradually become a task that has not yet been explored but has important practical significance. In this article, we propose a heuristic algorithm with budget elements on multi-layer networks to find the optimal combination of seed nodes under budget constraints. Through propagation dynamics experiments based on real and artificial networks, it has been proven that our proposed algorithm outperforms other benchmark algorithms in terms of influence propagation.
\end{abstract}

%%%%%%%%% BODY TEXT %%%%%%%%%%%%%%%%%%%%%%%%%%%%%%%%%%%%%%%%
\section{Introduction}
\label{s:introduction}
\noindent
How to increase the breadth and speed of information dissemination has garnered increasing attention among researchers due to its wide-ranging applications, such as promoting company products, mitigating negative impacts during crises, and driving social progress and innovation~\cite{1,2}. The increase in information diffusion breadth is generally referred to as the influence maximization (IM) problem \textendash a classical optimization problem that aims to select a fixed number of nodes to maximize influence spread through a specific diffusion mechanism~\cite{3,4,5,6}. In viral marketing, selecting a group of influential nodes on a social platform can help spread information about a brand or product widely, thus increasing purchase rates~\cite{7}. Furthermore, seeds derived from solutions to the IM problem under disease spread models can serve as an effective immunization strategy to control contagious diseases, such as COVID-19, H7N9 and AIDS~\cite{8,9,10}.

The classical IM problem only seeks optimal seeds to maximize their influence in a network, regardless of the cost of choosing them, resulting in algorithms hardly being applied to real scenarios~\cite{11,12}. For example, choosing celebrities on Sina Weibo as seeds to promote a product can significantly boost its visibility. However, such choices could lead to a high cost, limiting profits from sales~\cite{13,14}. Therefore, budgeted influence problem (BIM) has been further studied on single-layer social networks. For instance, Nguyen et al.~\cite{15} proposed an efficient greedy algorithm which is an approximation of the optimal solution with a certain ratio to the BIM problem, but the algorithm lacks scalability to large networks. Souza et al.~\cite{16} introduced a new seeding strategy called Node Surround to select influential nodes under the threshold diffusion model, which selects relatively low-cost nodes that are close to expensive and structurally privileged nodes
in a network. Zhang et al.~\cite{17} designed a meta-heuristic algorithm named IICEA based on the local-global influence indicator to solve the BIM problem. Additionally, Banerjee et al.~\cite{18} proposed a community-based solution approach, namely Combim, which selects the least expensive nodes from each community while ensuring the budget constraint is met.

% However, Combim is suitable for BIM without seed size constraints, and in reality, the number of seeds also represents a cost.
Even though the above-mentioned methods have successfully solved the BIM problem to some extent in single-layer social networks, there are still some challenges that need to be addressed~\cite{19,20}. Firstly, previous studies to solve the BIM problem only considered budget constraints, ignoring the number of seeds to be selected. This may result in the selection of a large fraction of nodes as seeds, which is impossible to achieve and also poses a burden in practice. Besides, users may have accounts on different platforms and can transmit information through these social platforms simultaneously~\cite{21,22,23,24,25,26,27,28,29}. Consequently, how to select seed nodes to maximize their influence on such multilayer networks needs to be studied. 

Therefore, we propose a new and more realistic IM problem, i.e., budget and capacity constrained influence maximization (BCIM) in multilayer networks, which considers both the budget and the size of the seed set upon the influence maximization problem. 
To solve the BCIM problem, we propose a multilayer multi-population genetic algorithm (MMGA) that takes into account both the topological properties of a multilayer network and the spreading ability of a node. Specifically, MMGA is inspired by biological evolution and consists mainly of three operators, i.e., crossover, mutation, and selection. In contrast to traditional genetic algorithms, MMGA incorporates a multi-population parallel evolution module to meet seed size constraint and accelerate the efficiency of the algorithm, and a repair module to address situations where the budget is exceeded during the evolution process. Comparative experiments performed on synthetic and empirical multilayer networks under the multilayer independent cascade model show that MMGA surpasses other state-of-the-art benchmarks in terms of influence spread under various constraints. Moreover, we further explore how different parameters, such as crossover probability, mutation probability, evolutionary step, and initialization mode, affect the performance of MMGA.

The remainder of this paper is structured as follows. In Section \textbf{2}, we provide the preliminary definition of a multilayer network and the spreading dynamics on it. In Section \textbf{3}, we show the detailed structure of MMGA and use examples to illustrate its key modules. We further illustrate the baselines extended from other IM problems and give detailed descriptions of the empirical multilayer networks that will be used in the subsequent experiment section in Section \textbf{4}.  Furthermore, the effectiveness of the proposed algorithm is evaluated in Section \textbf{5}. Finally, we conclude this work in Section \textbf{6}.

\section{Preliminary definition}
\label{sec:PreliminaryKnowledge}
\noindent
\subsection{Problem Definition}
\noindent
A multilayer network $G=\{G^1,\cdots,G^m,\cdots,G^M\}$ contains $M$ layers of networks, where the $m-$th layer is denoted as $G^m=(V,E^m,C)$. We assume that each layer contains the same set of nodes, which is denoted as $V=\{v_1,v_2,\cdots,v_N\}$. Each node may have different connections in different layers, so we use $E^m=\{e^m_1,e^m_2,\cdots,e^m_h\}$ to represent the edge set in the $m-$th layer. Multilayer networks are ubiquitous in real-world systems. For instance, on social platforms, users may have various connections representing different types of relationships, such as friendship, kinship, or co-workership~\cite{30}. By representing each platform as a separate layer, we can construct a multilayer network based on these diverse user connections. Additionally, a multilayer network can represent various types of transportation connections between cities, such as buses, trains, and airlines, with each layer corresponding to a specific mode of transportation~\cite{31}.

We focus on the budget and capacity constrained influence maximization problem (BCIM), in which the cost of inviting users to do an advertisement or disseminate a piece of information may differ according to the popularity of the users. Therefore, we associate each node with a cost, i.e.,  we use $C=\{c_1,c_2,\cdots,c_N\}$ to represent the cost of each node and $c_j=\sum^M_{m=1}d^m_j$, where $d^m_j$ is the degree of node $v_j$ in the network of $m-$th layer. This means that we assume that the cost of a node to be chosen as the seed is positively correlated with the sum of the degree of the node in the $M$ layers.

The conventional budgeted influence maximization (BIM) problem seeks to maximize influence spread within a budget by selecting seed nodes, but without limiting the number of seeds, which may lead to impractically large sets of low-cost nodes as solutions. For instance, in the context of viral marketing, it would be impractical for marketing entities to rely on numerous low-cost promoters (referred to as seed nodes in the IM problem), as this approach demands considerable time and carries the risk of ineffective promotion.
Therefore, we formulate a more practical influence maximization problem, i.e., a budget and capacity constrained influence maximization problem in multilayer networks (BCIM). Given a specific spreading model, the goal of BCIM is to select up to $K$ nodes whose total cost is with $B$ as seeds to maximize their influence in a multilayer network, which is mathematically given as follows

\begin{equation}
\left\{
\begin{array}{lr}
    {\operatorname{argmax}}_{S \subseteq V} |\customcup{\text{$m$=1}}{\text{M}}V^{m}_a(S)|, & \\
    s.t. \quad |S| \leq K , C(S)\leq B, \\
\end{array} 
\right.
\end{equation}
where $S$ denotes the seed set, and $V^{m}_a(S)$ is a set that contains nodes in $m$-th layer that are influences by the seed nodes in $S$. Moreover, $C(S)=\sum_{v_i \in S}c_i$ means the cost of seeds and $B$ is the predefined budget.

\subsection{Multilayer independent cascade model}
\noindent
To quantify node influence, we utilize the multilayer independent cascade model (MIC) to simulate the spreading process across different layers of a multilayer network. In MIC, nodes can be in one of two states, i.e. active or inactive. If a node is activated in one layer, it will immediately be active in all layers. An inactive node can be activated by an active node with probability $p$ and remain in the active state permanently. Remarkably, an active node could only activate its inactive neighbors once.
The specific details of MIC are as follows.
\begin{itemize}
\item Initially, the nodes in the seed set $S$ are assigned to the active state, and the remaining nodes are inactive. We use $S_{t-1}$ to represent the nodes that are newly activated at time step $t-1$, and thus we have $S_{0}=S$.

\item 
In step $t$, for each node $ v_i \in S_{t-1}$, we first find its inactive neighbors in all layers and name it $N_{ina}(v_i)$. We note that $N_{ina}(v_i)$ may contain repeated nodes $v_k$, indicating $v_k$ being the neighbor of $v_i$ at different layers. For every node $v_j \in N_{ina}(v_i)$, it will be activated by $v_i$ with probability $p$. The newly activated nodes at step $t$ are inserted into the set $S_{t}$.

% . The nodes are activated at time step $t-1$ represent as $A^{t-1}$, the inactive neighbors of node $v_i$ ($v_i \in A^{t-1}$) are denoted as $N^m_{ina}(v_i)$. For each $v_j \in N^m_{ina}(v_i)$, it will be activated by $v_i$ with independent probability $p$ and will be added into $A^{t}$ if activated. Therefore, set $A^{t}$ contains the nodes activated in step $t$ across all the layers.

\item The contagion process continues until there is no new node being activated.
\end{itemize}

\begin{figure*}[h]
  \centering
  \includegraphics[width=0.8\linewidth]{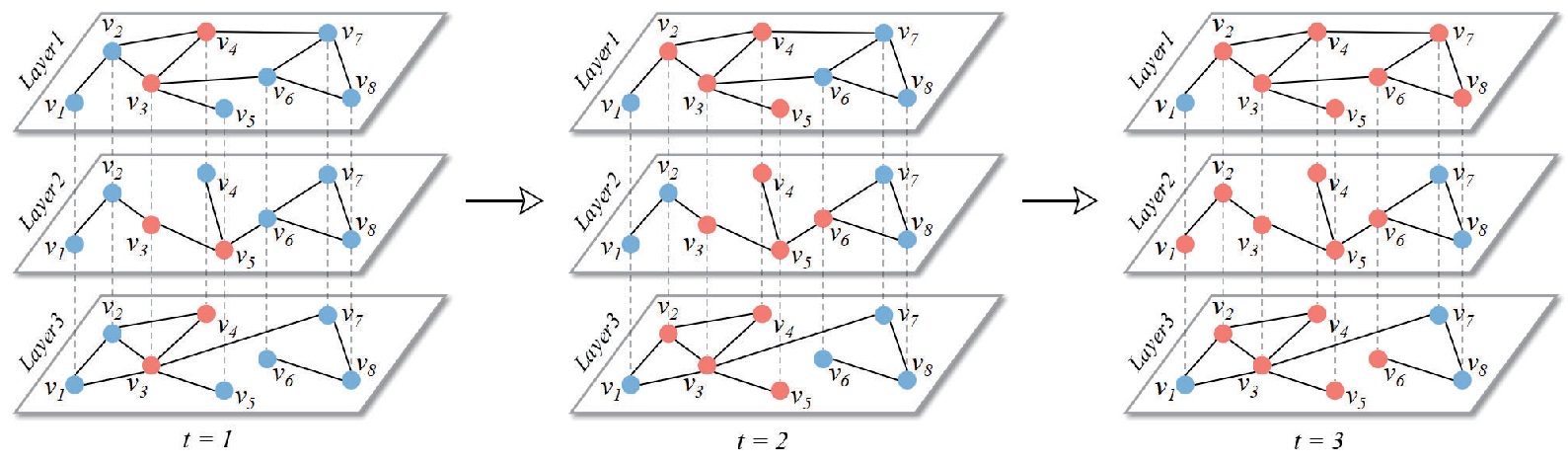}
  \caption{A schematic diagram of the MIC model in a three-layered network. We use red and blue to represent nodes in active and inactive states, respectively.}
  \label{MIC}
\end{figure*}

We show an example of the MIC model in Figure~\ref{MIC}, where the initial seed node is $v_3$. In step $t=1$, the set of inactive neighbors of $v_3$ in all layers is $N_{ina}(v_3)=\{v_1, v_2, v_2, v_2, v_4, v_4, v_5, v_5, v_5, v_6, v_7\}$. Every node in $N_{ina}(v_3)$ will be activated by $v_3$ with probability $p$ and thus $v_4$ and $v_5$ are activated and form the newly activated set $S_1=\{v_4, v_5\}$. Subsequently, each node in $S_1$ will activate their inactive neighbors with probability $p$ across three layers at time step $t=2$, respectively. At the end of step $t=2$, we get $S_2=\{v_2, v_6\}$. At time step $t=3$, the newly activated nodes of the three layers are $v_1$, $v_7$ and $v_8$. 

\begin{figure}[h]
  \centering
  \includegraphics[width=0.55\linewidth]{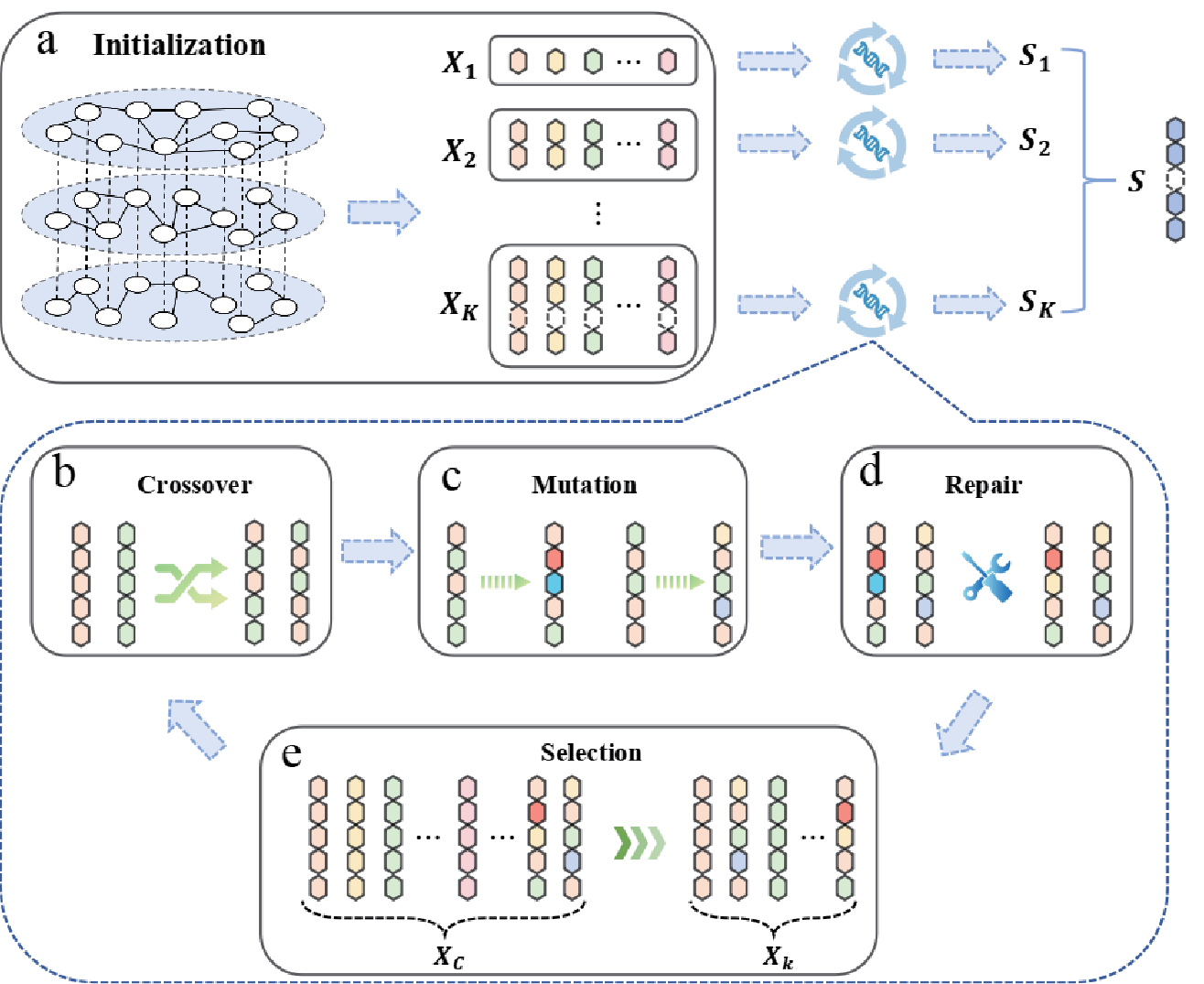}
  \caption{Framework of the multilayer multi-population genetic algorithm. \textbf{(a)} The initialization module generates $K$ populations. Each population contains $n$ individuals, which are selected from the multilayer network. \textbf{(b)} The crossover module exchanges certain nodes within individuals according to specific rules to generate new offspring. \textbf{(c)} The mutation module alters the nodes within the offspring generated in step (b) to randomly selected nodes. \textbf{(d)} The repair module aims to adjust the offspring that exceed the budget $B$. \textbf{(e)} The selection process retains the offspring with the higher fitness value as input for the next iteration.}
  \label{fig:Framework}
\end{figure}

\begin{algorithm}[h]
\caption{Multilayer Multi-population Genetic Algorithm  (MMGA)}\label{MMGA_pseudo_code}
  \SetKwInOut{Input}{Input}\SetKwInOut{Output}{Output}
  \Input{$G$, $B$, $K$, $p_c$, $p_m$, $T$}
  \Output{$S$}
  $O \leftarrow \emptyset$;\\
  \For{each $k \leq K$}{
  $X_k \leftarrow$ \textbf{Initialization}($G, V, k, n$);\\
  \For{each $x \in X_k$}{
   $x \leftarrow$ \textbf{Repair}($x, V, B$);\\
  }
  $x_{ko} \leftarrow$ randomly select from $X_k$;\\
  
  \For{each $t \leq T$}{
    $X_{os} \leftarrow$ \textbf{Crossover}($X_k$, $k$, $p_c$);\\
    $X_C \leftarrow$ \textbf{Mutation}($V$, $X_{os}$, $X_k$, $p_m$);\\
    \For{each $x \in X_C$}{
    $x \leftarrow$ \textbf{Repair}($x, V, B$);
    }
    $X_k \leftarrow \emptyset$;\\
    \While{$|X_k|<n$}{
    $x_i \leftarrow \operatorname{argmax}_{x\in X_C}\{f(x)\}$;\\
    $X_k \leftarrow X_k \cup x_i$;\\
    remove $x_i$ from $X_C$
    }
     $x_{ko} \leftarrow {\operatorname{argmax}}\{\sigma(x_{ki}), x_{ki} \in \{x_{ko}\}\cup X_k \}$;\\   
    }
    $O \leftarrow O \cup x_{ko}$;
  }
  $S \leftarrow {\operatorname{argmax}}\{\sigma(x), x \in O \}$

\end{algorithm}

\section{Multilayer Multi-population Genetic Algorithm (MMGA)}
\noindent
The introduction of two constraints in the BCIM problem may result in the loss of submodularity, making it difficult for a greedy algorithm—despite its high computational complexity—to find an optimal solution. While low-complexity heuristic algorithms might provide feasible solutions, they often do so with a reduced accuracy. To address this issue, we have developed a meta-heuristic algorithm known as the Multilayer Multi-Population Genetic Algorithm (MMGA), which emulates natural evolutionary mechanisms to search for the global optimal solution. To the best of our knowledge, this is the first time that multi-population evolution has been applied to the BCIM problem, which traditional genetic algorithms cannot effectively solve. Specifically, MMGA comprises initialization, repair and evolution modules, where each population evolves in parallel. We show the framework and pseudo-code of MMGA in Figure~\ref{fig:Framework} and Algorithm~\ref{MMGA_pseudo_code}-\ref{Mutation_code}. The details of each module in MMGA are given in the following subsections.

\subsection{Initialization}
\begin{figure*}[h]
  \centering
  \includegraphics[width=0.7\linewidth]{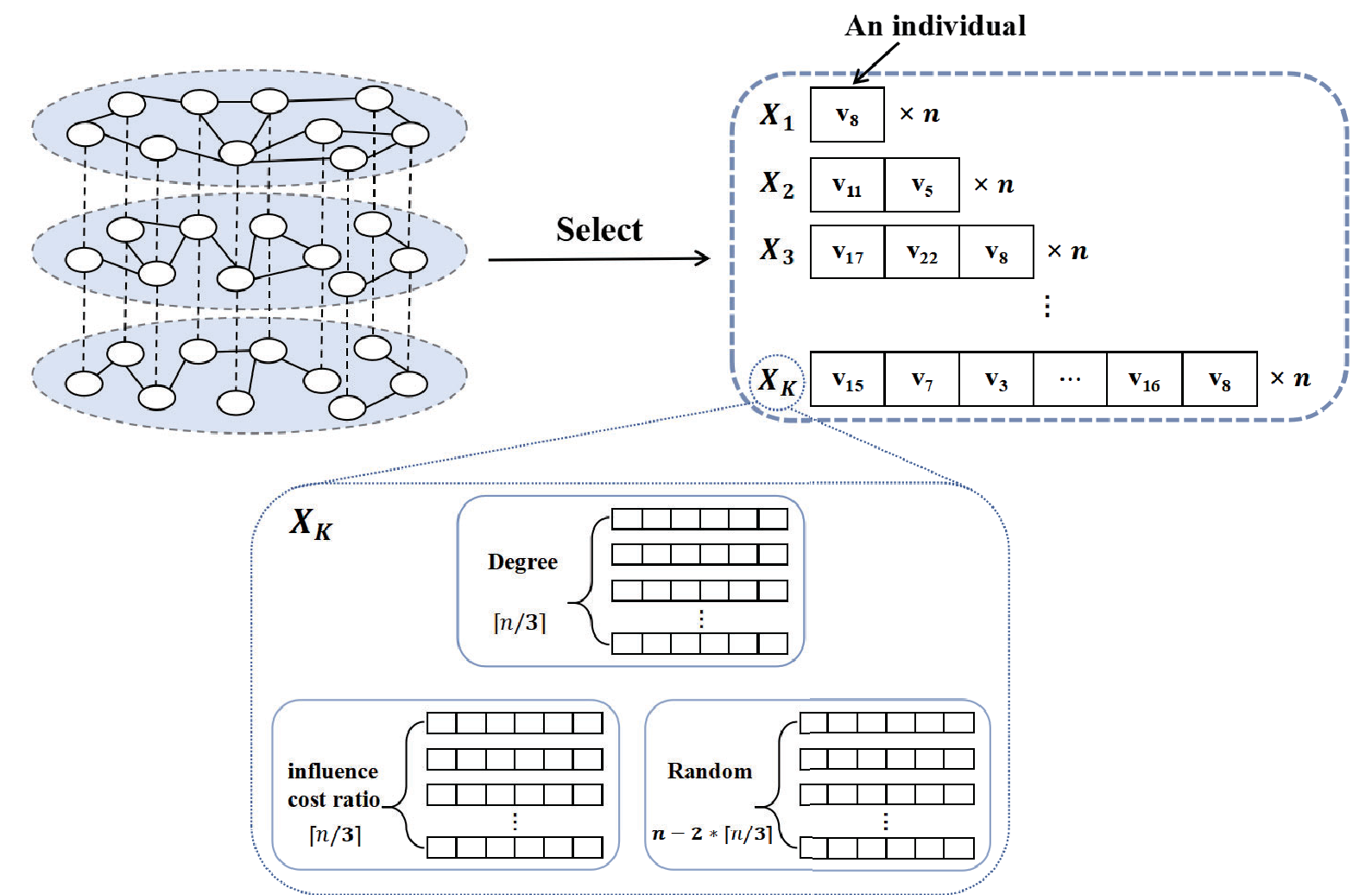}
  \caption{Initialization of the multilayer multi-population genetic algorithm.}
  \label{fig:Initialization}
\end{figure*}

\noindent
In the initial state, we generate a set that contains $K$ initial populations, i.e., $\{X_1, X_2,\cdots, X_K\}$.  Each population $X_k=\{x_{k1}, x_{k2}, \cdots, x_{kn}\}$ consists of $n$ individuals, and every individual is composed of a batch of nodes selected from the multilayer network as the initial solution for the BCIM problem. Specifically, for each individual $x_{ki}=\{v_{i_1},v_{i_2},\cdots,v_{i_k}\} \in X_k$, it contains $k$ non-duplicate nodes generated from $V$. As each $X_k$ contains $n$ individuals, we use three different methods, i.e, degree, influence-cost ratio, and random, to initialize the first $\left\lceil \frac{n}{3} \right\rceil$, the second $\left\lceil \frac{n}{3} \right\rceil$ and the last $n - 2\left\lceil \frac{n}{3} \right\rceil$ individuals in $X_k$ in order to accelerate the convergence of the algorithm.
The details of the methods are given below, and a more intuitive explanation is given in Figure~\ref{fig:Initialization}.

\textbf{i) Initialization with node degree.} 
For every individual $x_{ki} \in X_k$ $(i\in[1, \left\lceil \frac{n}{3} \right\rceil])$, we first choose the $k$ nodes with the highest degree as elements of $x_{ki}$. Then each node in $x_{ki}$ is replaced by a randomly chosen node in $V \backslash x_{ki}$ with probability $\beta$.

\textbf{ii) Initialization with influence-cost ratio (ICR).} The influence-cost ratio of nodes $v_i$ is defined as $\phi(v_i) = \frac{\sigma(v_i)}{c_i}$, where $\sigma(v_i)$ represents the actual spreading capability of node $v_i$, quantified by the final outbreak size in the population when selecting node $v_i$ as the seed, and $c_i$ denotes the cost of node $v_i$. For every individual $x_{ki} \in X_k$ $(i\in[\left\lceil \frac{n}{3} \right\rceil+1, 2\left\lceil \frac{n}{3} \right\rceil])$, we first choose the $k$ nodes with highest influence-cost ratio value as the elements of $x_{ki}$, and then each element in $x_{ki}$ is replaced with a node selected from $V\backslash x_{ki}$ with probability $\beta$.

\textbf{iii) Initialization randomly. } For every individual $x_{ki} \in X_k$ $(i\in[2\left\lceil \frac{n}{3} \right\rceil+1, n])$, the element are selected randomly from $V$ but should guarantee that there is no repeated nodes in $x_{ki}$.

\begin{algorithm}[h]
\caption{Initialization}\label{Initialization_pseudo_code}
\SetKwInOut{Input}{Input}\SetKwInOut{Output}{Output}
  \Input{$G$, $V$, $k$, $n$;}
  \Output{$X_k$, $x_{ko}$;}
   $X_k \leftarrow \emptyset$;
   
    \For{$i$ from $1$ to $\lceil n/3\rceil$}{
    $x_i \leftarrow$ max\_degree($G, k$);\\
        \For{$j$ from $1$ to $k$}{
         $p$ = random(0, 1);\\
         \textbf{if} $p \leq p_\alpha$ \textbf{then}:\\
            \quad $v^{'} \leftarrow$ randomly select from $V$;\\ 
            \quad $x_i[j] \leftarrow v^{'}$;\\ 
        }
        $X_k \leftarrow X_k \cup x_i$;\\
    }
    \For{$i$ from $1$ to $\lceil n/3\rceil$}{
    $x_i \leftarrow$ max\_ICR($G, k$);\\
        \For{$j$ from $1$ to $k$}{
         $p$ = random(0, 1);\\
         \textbf{if} $p \leq p_\alpha$ \textbf{then}:\\
            \quad $v^{'} \leftarrow$ randomly select from $G$; \\
            \quad $x_i[j] \leftarrow v^{'}$;\\ 
        }
        $X_k \leftarrow X_k \cup x_i$;\\
    }
    \For{$i$ from $1$ to $n-2*\lceil n/3\rceil$}{
    $x_i \leftarrow$ random\_selection($G, k$);\\
    $X_k \leftarrow X_k \cup x_i$;
    }
\end{algorithm}

\subsection{Repair Module}

\begin{figure*}[t]
  \centering
  \includegraphics[width=0.75\linewidth]{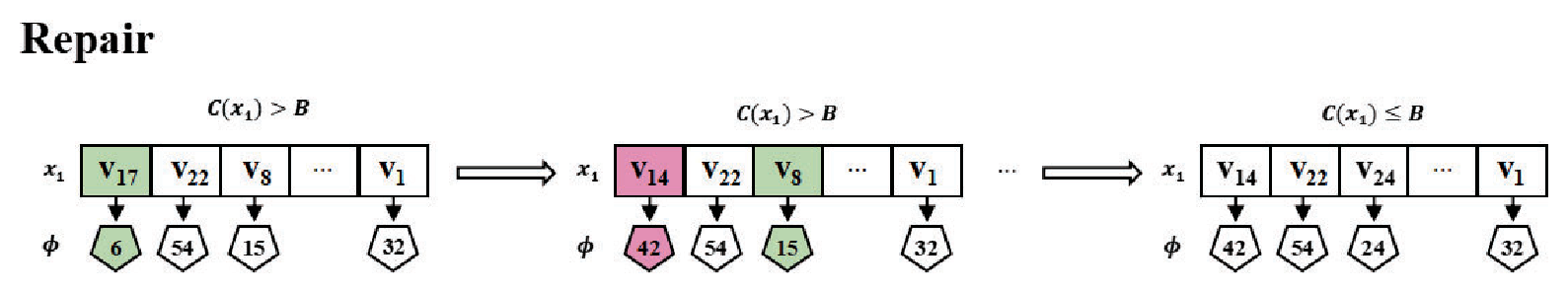}
  \caption{The repair module in the multi-population genetic algorithm.}
  \label{repair}
\end{figure*}

Since the populations initialized above may contain individuals that exceed the predefined budget $B$, we further design a repair module to adjust the individuals. The main idea of the repair module is to replace the low-influence-cost ratio score nodes with high-influence-cost ratio score nodes, thereby reducing the individual's cost while ensuring the influence spread. Given an individual $x_i$ that exceeds budgets, i.e., $C(x_i)>B$, we find the node that has the lowest influence-cost ratio score in $x_i$ and find a node that has the highest influence-cost in $V\backslash  x_i$ to replace it. The process ends at $C(x_i)\leq B$. We show an example of repair operation in Figure~\ref{repair}, where the individual $x_1$ exceeds the budgets. The node $v_{17}$ in $x_1$ has the lowest influence-cost ratio value $\phi(v_{17})=6$, so it is replaced by the node $v_{14}$ with $\phi(v_{14})=42$. But we still have $C(x_1)>B$. So, node $v_8$ with the lowest influence-cost ratio $\phi(v_{8})=15$ is further replaced by node $v_{24}$. 
We give the pseudo-code of repair module in Algorithm~\ref{repair_pseudo_code}.

\begin{algorithm}[h]
\caption{Repair}\label{repair_pseudo_code}
\SetKwInOut{Input}{Input}\SetKwInOut{Output}{Output}
  \Input{$x$, $V$, $B$;}
  \Output{$x$;}
    \While{$C(x)>B$}{
    index $\leftarrow$ index of node with the lowest \textbf{ICR} in $x$;\\
    $v^{'} \leftarrow$ node with the highest \textbf{ICR} in $V\backslash x$;\\
    $x[index] \leftarrow v^{'}$;\\
    }

\end{algorithm}

\subsection{Evolution}
\noindent
To accelerate the computation speed of the algorithm, we
apply a parallel evolution strategy, which includes crossover, mutation, and selection, to the initial population set, allowing each population to evolve independently and select its optimal solution. Taking $X_k (k=1,2, \cdots, K)$ as an example, we randomly choose an individual $x_{ko} \in X_k$ as the potential optimal solution in $X_k$. Then we update $x_{ko}$ using the evolution strategy for $T$ rounds to obtain the final optimal solution for $X_k$. Therefore, we can get $K$ optimal solutions for the $K$ populations, i.e., $O=\{x_{1o}, x_{2o}, \cdots, x_{Ko}\}$, using the parallel evolution strategy.
Finally, we choose the individual with the highest spread of influence from $O$ as the final optimal solution for BCIM. Algorithm~\ref{corssover_code} and \ref{Mutation_code} show the specific pseudo-code of crossover and mutation. Moreover, we show the details of the evolution strategy using $X_k (k=1,2, \cdots, K)$ as an example as follows.

\begin{itemize}
\item \textbf{Crossover:} The crossover operator generates various combinations of nodes, offering a wide range of potential solutions for the combinatorial optimization problem. Given population $X_k$, we randomly choose two individuals from it, namely $x_{kp}=\{v_{p_1},v_{p_2},\cdots,v_{p_k}\}$ and $x_{kq}=\{v_{q_1},v_{q_2},\cdots,v_{q_k}\}$. Each corresponding position in $x_{kp}$ and $x_{kq}$ is exchanged with probability $p_c$ and under the condition that no duplicate nodes exist within the resulting offspring individuals, i.e., $x_{kp}^{'}=\{v_{p_1}^{'},v_{p_2}^{'},\cdots,v_{p_k}^{'}\}$ and $x_{kq}^{'}=\{v_{q_1}^{'},v_{q_2}^{'},\cdots,v_{q_k}^{'}\}$. The above process is repeated for $R$ times, resulting in $2R$ new offspring. 

\begin{algorithm}[h]
\caption{Crossover}\label{corssover_code}
\SetKwInOut{Input}{Input}\SetKwInOut{Output}{Output}
  \Input{$X_k$, $k$, $p_c$;}
  \Output{New offspring $X_{os}$;}
  $X_{os} \leftarrow \emptyset$;\\
  \For{$i$ from $1$ to $R$}{
  $x_{kp}^{'}, x_{kq}^{'} \leftarrow$ randomly select from $X_k$;\\
  \For{$j$ form $1$ to $k$}{
  $p$ = random(0, 1);\\
  \textbf{if} $p \leq p_c$ and $x_{kp}^{'}[j] \notin x_{kq}^{'}$ and  $x_{kq}^{'}[j] \notin x_{kp}^{'}$ \textbf{then}:\\
  \quad switch $x_{kp}^{'}[j]$ and $x_{kq}^{'}[j]$;\\
  }
  $X_{os} \leftarrow X_{os} \cup x_{kp}^{'}$;\\
  $X_{os} \leftarrow X_{os} \cup x_{kq}^{'}$;
  }
\end{algorithm}

\item \textbf{Mutation:} To further expand the solution space, the MMGA algorithm performs a mutation operator on the offspring individuals generated by the crossover operator. In particular, each node in the offspring $x_i$ is replaced by a random node chosen from $V \backslash x_i$ with a probability of $p_m$. It should be noted that the $2R$ new offspring generated by crossover and mutation operators may exceed the budget. Hence, we further use the repair module illustrated in the initialization step to further repair the offspring. Finally, a new population $X_C$ is generated, which contains $n+2R$ individuals composed of individuals of the offspring $X_k$ and $2R$ generated by the crossover and mutation operator.

For the sake of clarity, we give detailed examples to further visualize the crossover and mutation operators, which are shown in Figure~\ref{Crossover_Mutation}. We use a crossover operator to exchange every position in $x_1$ and $x_2$. In the first position, we generate a random number $y$, and since $y$ is smaller than $p_c$, $v_8$ and $v_{12}$ are exchanged. However, a random number $y$ is generated for the node pairs $v_6$ and $v_5$ at the second position, they are not swapped since $y\geq p_c$. Subsequently, even though $y$ is smaller than $p_c$ in the third position for the pair of nodes $v_{11}$ and $v_6$, they cannot exchange as $v_6$ is in $x_1$ in the second position. The crossover operator will work on all the positions in $x_1$ and $x_2$, and finally we can obtain two offspring $x_1^{'}=\{v_{12},v_{6},v_{11},\cdots,v_{3}\}$ and $x_2^{'}=\{v_{8},v_{5},v_{6},\cdots,v_{2}\}$.
We use offspring $x_1^{'}$ to illustrate the mutation process, in which each node is mutated with probability $p_m$. We generate a random number $y$ for each node in $x_1^{'}$ and only mutate the nodes when $y<p_m$. For example, node $v_6$ is replaced by $v_{16}$, which is chosen randomly from $V \backslash x_1^{'}$.

\begin{figure*}[t]
  \centering
  \includegraphics[width=0.8\linewidth]{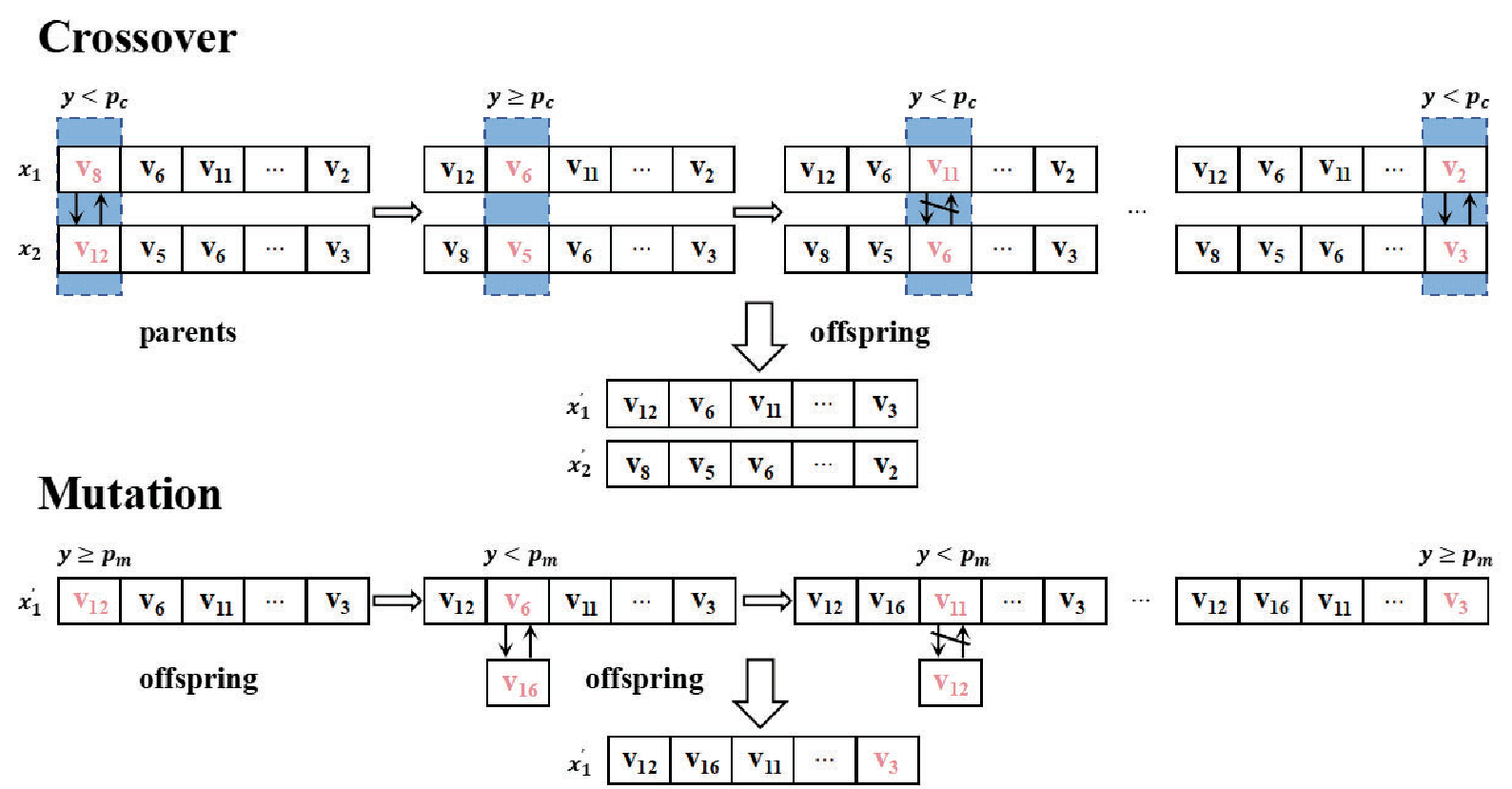}
  \caption{The evolution of multi-population genetic algorithm.}
  \label{Crossover_Mutation}
\end{figure*}

\begin{algorithm}[h]
\caption{Mutation}\label{Mutation_code}
\SetKwInOut{Input}{Input}\SetKwInOut{Output}{Output}
  \Input{$V$, $X_{os}$, $X_k$, $p_m$;}
  \Output{$X_C$;}
  $X_C \leftarrow X_k$;\\
  \For{$i$ from $1$ to $2*R$}{
   \For{$j$ form $1$ to $k$}{
   $p$ = random(0, 1);\\
   \textbf{if} $p \leq p_m$ \textbf{then}:\\
   \quad $v^{'} \leftarrow$ randomly select from $V\backslash X_{os}[i]$;\\
   \quad $X_{os}[i][j] \leftarrow v^{'}$;\\
   }
   $X_C \leftarrow X_C \cup X_{os}[i]$;
  }
\end{algorithm}

\item \textbf{Selection: }The selection module aims to select $n$ individuals with high fitness scores from $X_C$ as the new population $X_k$ for the next iteration. For an individual $x_i$, the fitness score is defined as
% \begin{equation}\label{equ:ICR}
    $f(x_i) = \sigma(x_i)$,
% \end{equation}
where $\sigma(x_i)$ represents the final influence spread of $x_i$ when it is chosen as the seed set under the MIC model. We use the roulette wheel algorithm to select $n$ individuals from $X_C$ to form a new population $X_k$. Particularly, the probability that $x_i$ is selected is $f(x_i)/\sum_{x_j \in X_C} f(x_j)$, meaning that individuals with high fitness scores have a high chance to be chosen. After we get the updated $X_k$, we find the optimal individual in it, i.e., $x_{ko}^{'}={\operatorname{argmax}} \{\sigma(x_{ki}), x_{ki} \in X_k\}$. Then, we update $x_{ko}$ by

\begin{equation}
% \left\{
% \[
x_{ko} =
\begin{cases}
    x_{ko}^{'}, & \text{if } x_{ko}^{'}>x_{ko}\\
    x_{ko}, & \text{if } otherwise \\
\end{cases}
% \]
% \right.
\end{equation}

\end{itemize}

In summary, MMGA algorithm initializes multiple populations with prior knowledge through its initialization module to promote rapid convergence. Each population then independently seeks a local optimal solution in parallel within the evolution module. We have also designed a repair function to handle cases where individuals might exceed the budget during the evolution process. Finally, the global optimal solution is selected from the local optimal solutions of each population, forming the final set of seeds.

\section{Baselines and Datasets}
\noindent
\subsection{Baselines}
\noindent
To validate the effectiveness of MMGA in solving the BCIM problem, we extend a series of baseline methods, which were originally designed for the conventional influence maximization problem or budgeted influence maximization problem, to solve the BCIM problem. We choose state-of-the-art algorithms, including meta-heuristic algorithms, such as DPSO, Greedy algorithm, and heuristic algorithms, i.e., ACD, Degree, and Community-based solution approach as baselines.  DPSO is a representative optimization algorithm frequently used in combinatorial optimization problems. The greedy algorithm has been rigorously proven to achieve an approximation ratio of 1-1/e for the conventional IM problem, making it a challenging benchmark for researchers to surpass. Additionally, ACD is an IM algorithm tailored for multilayer network structures, while Combim is specifically designed for the budgeted influence maximization problem. The details of each algorithm are given below.

\textbf {Discrete Particle Swarm Optimization (DPSO)}~\cite{32}: We generalize the DPSO algorithm, which is inspired by the self-organizing behavior of bird flocks, from single-layer network to multilayer network as a baseline method. To solve the BCIM problem, the algorithm generates $n$ particles as $n$ possible solutions. The $n$ particles evolve simultaneously and the one with the highest fitness score, which is quantified by the final influence spread of the seeds by  Monte Carlo simulations, will be chosen as the optimal seed set. Each particle $i$ is associated with a position vector $\boldsymbol{x_i}=(v_{i1},v_{i2},\cdots,v_{iK})$ which is initialized by degree, with non-duplicate nodes, and a velocity vector $\boldsymbol{V_i}=(r_{i1},r_{i2},\cdots,r_{iK})(r_{ik} \in \{0, 1\})$ which is initialized by all zero values.

The velocity vector $\boldsymbol{V_i}$ determines the optimization direction of the particle $i$ and is updated by the best personal position vector $\boldsymbol{x_i^{Pbest}}=(v_{i1}^{Pbest},v_{i2}^{Pbest},\cdots,v_{iK}^{Pbest})$ and the best global position vector $\boldsymbol{x^{Gbest}}=(v_{1}^{Gbest},v_{2}^{Gbest},\cdots,v_{K}^{Gbest})$. At time step $t$, for particle $i$,  if its current fitness score is larger than the historical optimum, the current particle's position vector becomes its best personal position $\boldsymbol{x_i^{Pbest}}$. 
The position vector of a particle with the highest fitness score in the swarm, i.e., among all particles, is considered the best global solution $\boldsymbol{x^{Gbest}}$ at the time step $t$. The velocity vector of the particle $i$ is updated via the following formula
\begin{equation}
    \boldsymbol{V_i^{'}} = \textbf{H}(w\boldsymbol{V_i} + I_1 J_1 (\boldsymbol{x_i} \cap \boldsymbol{x_i^{Pbest}}) + I_2 J_2 (\boldsymbol{x_i} \cap \boldsymbol{x^{Gbest}})),
    \label{V_update}
\end{equation}
where $w$ represents the inertia weight, parameters $I_1$ and $I_2$ are learn factors, $J_1$ and $J_2 \in [0,1]$ denote random numbers. The operator $\cap$ is a similar intersection operation, in which if $v_{ik}$ of $\boldsymbol{x_i}$ is also in $\boldsymbol{x_i^{Pbest}}$, then the $k$-th position in $\boldsymbol{x_i} \cap \boldsymbol{x_i^{Pbest}}$ is set to $0$; otherwise, it is set as 1. In addition,  $\textbf{H}(\cdot)$ is defined as a threshold function shown as follows

\begin{equation}
\textbf{H}(q)
\left\{
\begin{array}{lr}
    0, \quad if \quad q<2& \\
    1, \quad if \quad q\geq 2\\
\end{array} 
\right.
\end{equation}
Therefore, we can further update particle $i$ via the updated velocity vector $\boldsymbol{V_i^{'}}$ using the following formula

\begin{equation}
\boldsymbol{x_i} \bigoplus \boldsymbol{V_i^{'}} = x^{'}_i = \{v^{'}_{i1},v^{'}_{i2},\cdots,v^{'}_{iK}\} 
\end{equation}

\begin{equation}
% \left\{
% \[
v^{'}_{ik} =
\begin{cases}
    v_{ik}, & \text{if } r_{ik}^{'}=0 \\
    replace(v_{ik}, V \backslash x_i), & \text{if } r_{ik}^{'}=1
\end{cases}
% \]
% \right.
\end{equation}
The operator $\bigoplus$ drives the particles to the promising regions, and the function of $replace$ means selecting a random node from $V \backslash x_i$ to replace $v_{ik}$. The best global solution for the BCIM problem is obtained after $T$ iterations of the above procedures.

\textbf {Greedy}: We propose a greedy algorithm to solve the BCIM problem, in which the influence spread of a node is quantified by Monte Carlo simulation. First of all, the node with the highest influence spread is selected as the first seed. Then we continue the process by iteratively adding nodes with the highest marginal influence gain into the seed sets until the the predetermined size of the seed set is reached or the budget is exhausted.

\textbf {Adaptive Coupling Degree (ACD)~\cite{33}} algorithm considers the adaptive coupling degree to find seed nodes for BCIM problem. 
The number of non-repeated neighbors across all the layers of a node $v_i$ is assigned as the initial value of the adaptive degree of $v_i$. In each iteration, ACD selects the node with the highest adaptive coupling degree as the seed node and update the adaptive degree of each neighbor $v_q$ of the corresponding seed node using the following formula

\begin{equation}
    D_c^{'}(v_q) = D_c(v_q)-e^{Q(v_q)},
\end{equation}
where $Q(v_q)$ represents the number of seed nodes in the neighborhood of node $v_q$. The iterative selection process continues until the number of seed nodes reaches $K$ or the budget is exhausted.

\textbf {Degree} assumes that the nodes with more neighbors are more influential in a multilayer network. For node $v_i$, we use the sum of the node degree across different layers to quantify its spreading ability, and then the nodes are sorted in a descending order based on their spreading ability. Subsequently, we select the nodes with top spreading ability as the seeds until the size of the seed set reaches $K$ or the budget is exhausted.

\textbf {Community-based solution approach for the BCIM problem (ComBim)~\cite{18}} addresses the BCIM problem based on community structure, which uses the community detection algorithm proposed by Pan et al.~\cite{34} to detect communities in all layers of a multilayer network. In the community detection algorithm, the similarity between two nodes $v_i$ and $v_j$ is defined as $Sim_{ij}^m=\sum_{v_z\in N^m(v_i) \cap N^m(v_j)} \frac{1}{d^m_z}$, where $N^m(v_z)$ and $d^m_z$ represent the neighboring set and the degree of node $v_z$ in the $m$-th layer, respectively. After obtaining all communities in the multilayer network, we allocate budgets to these communities according to their size, i.e., communities with larger sizes can get more budgets and vice versa. The selection of seed nodes starts from the smallest community until the $K$ seed nodes are selected or the budget is exceeded. In each of the communities, we select nodes with high degree as the seeds until the budget of this community is exceeded.
The remaining budget of a community is transferred to the community with the most connections to it. The selection process continues until the number of seed nodes reaches $K$ or the total budget is exhausted.

\subsection{Empirical and synthetic multilayer networks}
We summarize three real-world multilayer networks and three artificial multilayer networks to evaluate the effectiveness of the algorithms discussed in the subsequent sections. The basic topological characteristics of these multilayer networks are presented in Table~\ref{tab:1}. The empirical multilayer networks are derived from transportation systems, social networks, and biological systems, respectively, and have been widely utilized in studies of resilience analysis, spreading dynamics, and structural reducibility, among other applications~\cite{35,36,37}.

\subsubsection{Empirical multilayer networks}

\noindent \hspace{1.3em} \textbf {London Multiplex Transport Network (Transport).~\cite{38}} The data was collected in 2013, which uses edges to represent whether there is a route between two stations (i.e., nodes in the network) or not. It contains three layers, illustrating stations connected by underground line (U), overground (O) and Docklands light railway (D), respectively.

\textbf {CKM Physicians Innovation Network (CKM).~\cite{39}}  The edges in this multilayer network were collected among physicians in four towns in Illinois, i.e., Peoria, Bloomington, Quincy and Galesburg. Each layer is constructed based on the each of the following questions:\\
 i) Who do you typically consult when you need advice about therapy?\\
ii) Who are the three or four physicians with whom you generally discuss cases or treatments during a typical week, such as last week?\\
iii) Could you provide the first names of the three or four people you spend the most social time with?\\
The three layers are named shortly Q.1, Q.2, and Q.3 in the following study, respectively.

\textbf {Drosophila Multiplex Gpi Network (Drosophila).~\cite{40}} In this data, we use edges to represent the interactions between genetics and proteins. The interactions in the three layers are direct interactions (DI), suppressive interaction (SI), and Physical association (Phy), respectively.

\subsubsection{Synthetic multilayer networks}

\noindent \hspace{1.3em} \textbf{ER Multilayer Network.} The construction of a single-layer ER network starts with a graph with $N=1000$ isolated nodes. For each pair of nodes $(v_i, v_j)$ where $v_i \neq v_j$, we add an edge between them with a probability of $0.004$. We generate three $ER$ networks with the same set of nodes to form a three-layer ER network.

\textbf{WS Multilayer Network.} The single-layer WS network starts with a ring structure of $1000$ nodes, with each node connecting to its four nearest neighbors. To introduce randomness, we rewire each edge with a probability of $0.3$. This process is repeated to create three WS networks with the same set of nodes and finally form a three-layer WS network.

\textbf{BA Multilayer Network.} To construct the BA network, we start with three initial nodes connected by two edges. At each time step, a new node is introduced and connects to two existing nodes using a preferential attachment mechanism, i.e., connecting to nodes with high degree. The process continues until the network reaches $1000$ nodes. We use the same node set to create three BA networks to form a three-layer BA network.

\begin{table}[htbp]%调节图片位置，h：浮动；t：顶部；b:底部；p：当前位置
	\centering
	\caption{Topological characteristics of the empirical multilayer networks, in which we use $N$ and $|E|$ to represent the number of nodes and edges of each layer, $<d>$ denotes the average degree of nodes, $<l>$ is the average shortest path, $L$ represents the diameter, $C$ represents the clustering coeffcient and $D$ represents the link density of each layer.\\}
    
	\label{tab:1} 
    \resizebox{0.7\linewidth}{!}{
	\begin{tabular}{ccccc cccc}%表格中的数据居中，c的个数为表格的列数
		\hline\hline\noalign{\smallskip}	
		Dataset& Layer & $N$ & $|E|$ & $<d>$ & $<l>$ &$L$& $C$ & $D$\\
		\noalign{\smallskip}\hline\noalign{\smallskip}
		 & U & 271 & 312 & 2.3& 13.96 & 39 & 0.031 & 0.0085 \\
		Transport& O & 83 & 83 & 2.0& 13.49 & 35 & 0.0 & 0.0244 \\
         & D & 45 & 46 & 2.04& 8.23 & 23 & 0.019 & 0.0465 \\
         \noalign{\smallskip}\hline\noalign{\smallskip}
         & Q.1 & 215 & 449 & 4.18& 2.90 & 7 & 0.260 & 0.0195\\
		CKM& Q.2 & 231 & 498 & 4.31& 2.94 & 8 & 0.260 & 0.0187\\
         & Q.3 & 228 & 423 & 3.71& 3.45 & 9 & 0.211 & 0.0163\\
        \noalign{\smallskip}\hline\noalign{\smallskip}
        &DI & 3126 & 5472 & 3.50 & 0.99 & 13 & 0.024 & 0.0011\\
        Drosophila & SI & 239 & 270 & 2.26& 1.30 & 7 & 0.174 &0.0095\\ 
        &Phy & 120 & 160 & 2.67& 1.41 & 9 & 0.016 & 0.0224\\
        \noalign{\smallskip}\hline\noalign{\smallskip}
        &ER.1 & 1000 & 1974 & 3.95& 0.38 & 11 & 0.004 & 0.0040\\
        ER &ER.2 & 1000 & 1996 & 3.99& 0.37 & 11 & 0.005 & 0.0040\\ 
        &ER.3 & 1000 & 1994 & 3.99& 0.34 & 11 & 0.002 & 0.0040\\
        \noalign{\smallskip}\hline\noalign{\smallskip}
        &WS.1 & 1000 & 2000 & 4.0& 6.13 & 11 & 0.179 & 0.0040\\
        WS &WS.2 & 1000 & 2000 & 4.0& 6.19 & 11 & 0.180 & 0.0040\\ 
        &WS.3 & 1000 & 2000 & 4.0& 6.23 & 11 & 0.199 & 0.0040\\
        \noalign{\smallskip}\hline\noalign{\smallskip}
        &BA.1 & 1000 & 1996 & 3.99& 4.03 & 7 & 0.026 & 0.0040\\
        BA &BA.2 & 1000 & 1996 & 3.99& 3.96 & 7 & 0.042 & 0.0040\\ 
        &BA.3 & 1000 & 1996 & 3.99& 4.13 & 8 & 0.017 & 0.0040\\
		\noalign{\smallskip}\hline
	\end{tabular}
    }
\end{table}

\section{Experiment}
\noindent
To validate the effectiveness of the MMGA algorithm, we conduct experiments on the three empirical multilayer networks and three types of synthetic multilayer networks. All experiments are carried out under the MIC model with a propagation probability $p=0.1$ unless otherwise specified, and the influence spread of a seed set is quantified by the average of the final outbreak size of $1000$ Monte Carlo simulations. To assess the effectiveness of the proposed method, we change the value of the seed size $K$
from $10$ to $20$ with an interval of $2$, and the value of the budget $B$ from $200$
to $600$ with an interval of $50$.
All the algorithms are implemented in Python and run independently on a server with a 2.20GHz Intel(R) Xeon(R) Silver 4114 CPU and 90GB of memory.

\begin{figure*}[t]
\centering
    \includegraphics[width=0.75\linewidth]{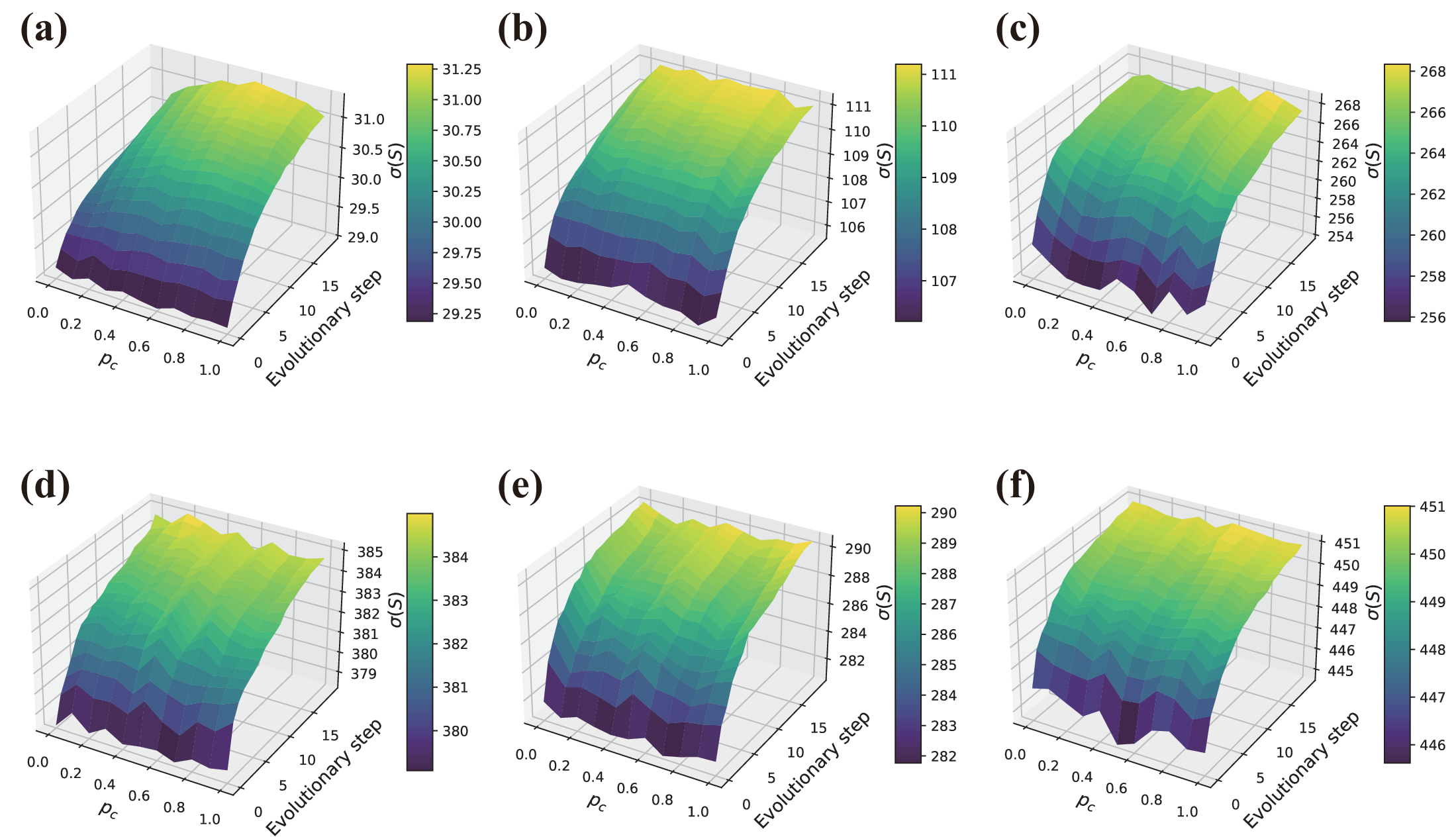}
% \captionsetup{font=small} % 设置字体大小为小号
 \caption{Performance of MMGA with the change of crossover probability $p_c$ and evolutionary step. We set $p_m=0.005$, $K=20$, and $B=600$. The results are given for networks: (a) Transport; (b) CKM; (c) Drosophila; (d) ER; (e) WS; and (f) BA.}
\label{fig:3D}
\end{figure*}

\subsection{Parametric analysis}
\begin{figure}[h]
\centering
    \includegraphics[width=0.5\linewidth]{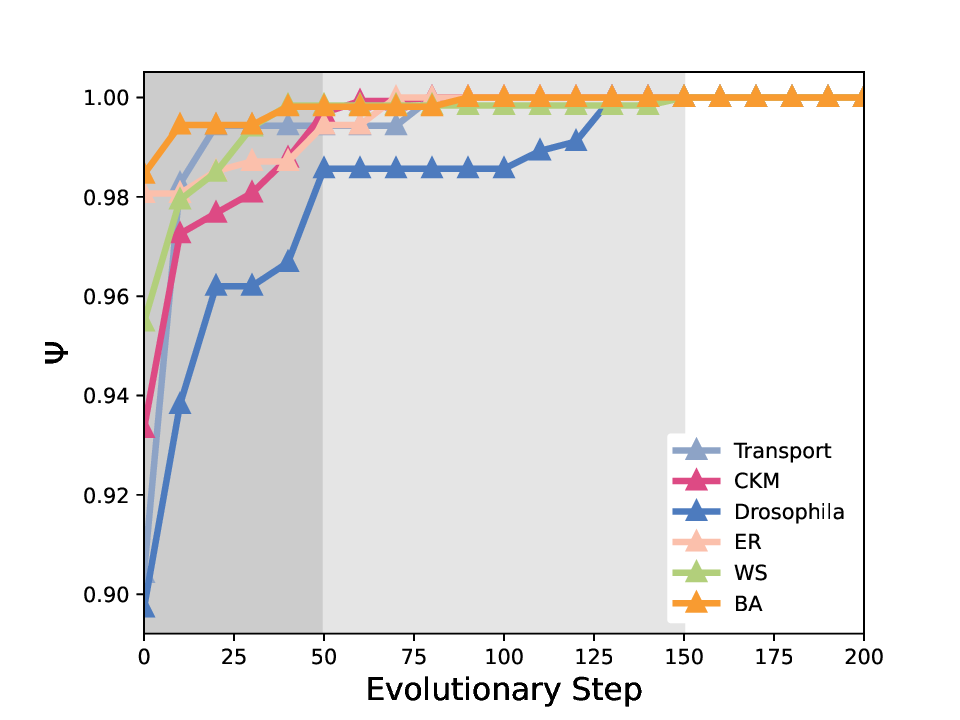}
 % \captionsetup{font=small} % 设置字体大小为小号
 \caption{The normalized influence spread $\Psi$ of MMGA when the evolutionary step changes. We set $p_c=0.8$, $p_m=0.005$, $K=20$, and $B=600$. The dark gray and light gray areas represent the rapid and gradual improvement phases of algorithm performance, respectively.}
 \label{fig: Evolutionary_Step}
\end{figure}

\noindent
In MMGA, parameters such as the probability of crossover $p_c$, the evolutionary step, and the probability of mutation $p_m$ would affect the influence of the spread of seeds. Therefore, we first explore the impact of different crossover probabilities $p_c$ on the influence spread of seeds when the evolutionary time step changes in both empirical and synthetic multilayer networks. The remaining parameters are set as follows: mutation probability $p_m=0.005$, budget $B=600$, and seed set $K=20$. As shown in Figure~\ref{fig:3D}, the spread of influence fluctuates with an increase of $p_c$ for different values of evolutionary steps and networks. The high value of the spread of the influence is achieved when $p_c$ is in the range of $0.6$ to $0.9$. We also show how the influence spread changes with the change of the evolutionary step in Figure~\ref{fig:3D} and Figure~\ref{fig: Evolutionary_Step}. For clarity, we use the normalized influence spread (i.e., normalized by network size) as the y-axis in Figure~\ref{fig: Evolutionary_Step}, and the probabilities of crossover and mutation are set to $0.8$ and $0.005$, the constraints of budget and seed set size are set to $600$ and $20$, respectively. As shown in the two figures, the spread of influence increases dramatically when the evolutionary step increases. More concretely, when the evolutionary step is less than $50$, the performance of MMGA improves rapidly. In most multilayer networks, the algorithm performance enters a gradual improvement phase between $50$ and $150$ steps, and after the $150$ steps it becomes stable. Consequently, we use the probability of crossover $p_c$ and the evolutionary step as $0.8$ and $150$ in subsequent experiments, respectively.

\begin{figure}[h]
\centering
    \includegraphics[width=0.5\linewidth]{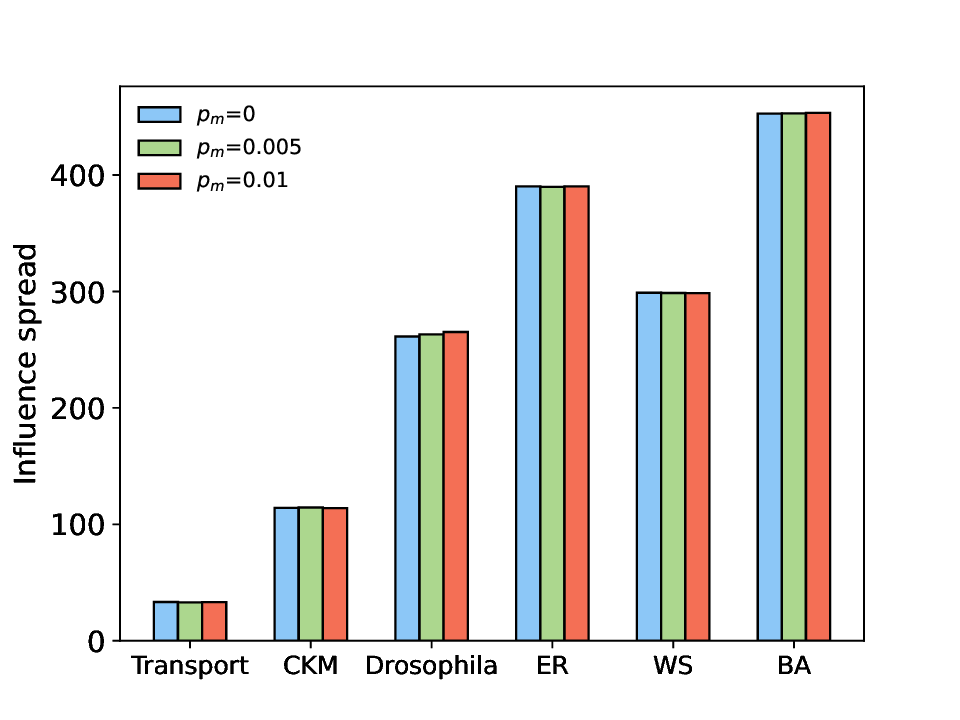}
 % \captionsetup{font=small} % 设置字体大小为小号
 \caption{The average influence spread of the seeds selected by MMGA under different mutation probability $p_m$. We set $p_c=0.8$, $K=20$, $B=600$, and $T=500$.}
 \label{fig: parameter_pm}
\end{figure}

In previous studies, the probability of mutation $p_m$ was generally low. We test the performance of the algorithm setting $p_m$ as $0$, $0.005$, and $0.01$, where the results are given in Figure~\ref{fig: parameter_pm}. We observe that the performance of MMGA with $p_m = 0.005$ and $0.01$ is slightly better than that of $p_m = 0$. And the performance of $p_m = 0.005$ and $0.01$ is similar in different networks. Therefore, we choose $p_m = 0.005$ in the following experiment.

\subsection{Algorithm performance}
\begin{figure*}[h]
\centering
    \includegraphics[width=0.8\linewidth]{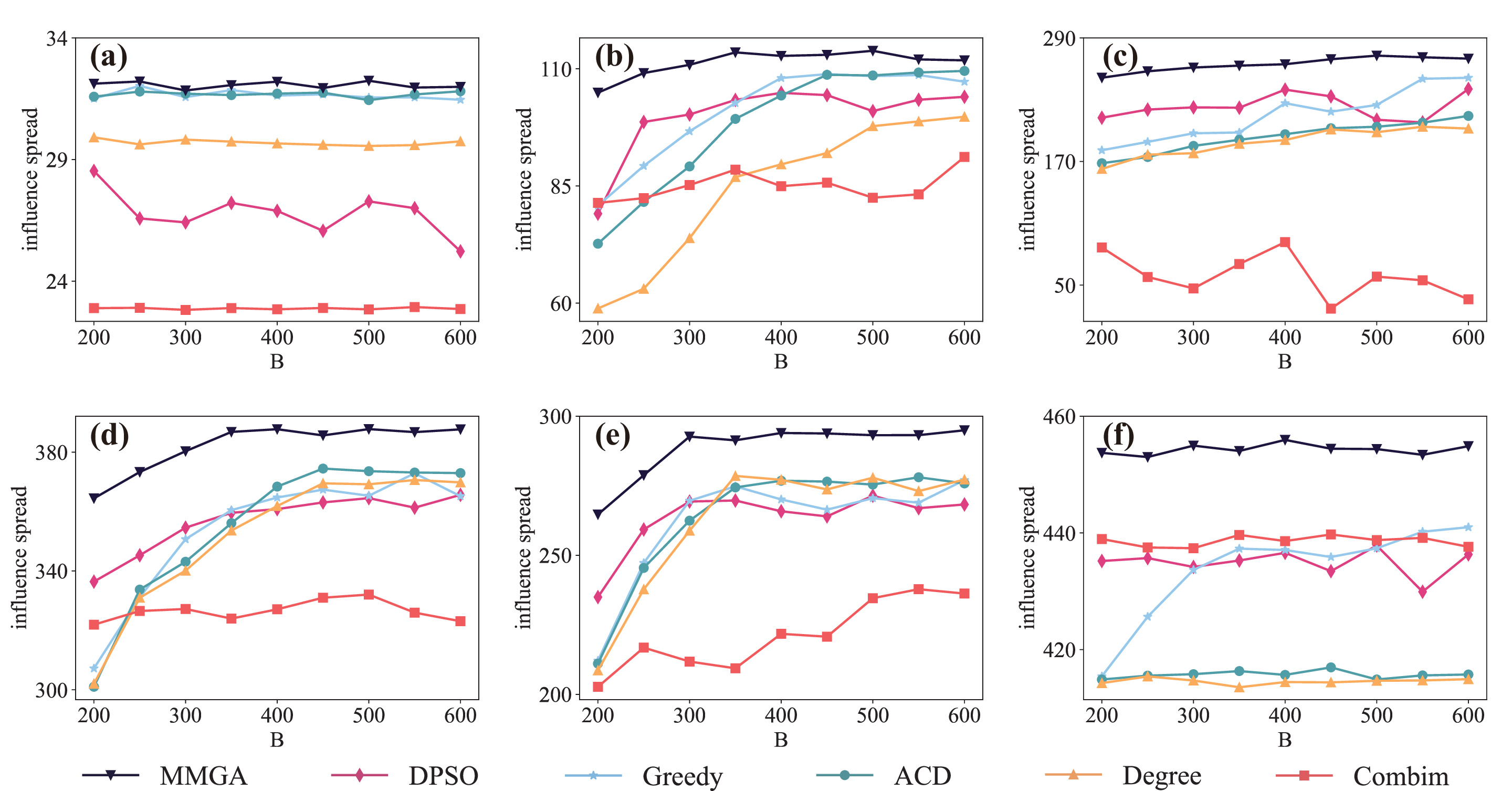}
 % \captionsetup{font=small} % 设置字体大小为小号
 \caption{The influence spread of different algorithms under various budget conditions, in which we set $K=20$. The results are given for networks: (a) Transport; (b) CKM; (c) Drosophila; (d) ER; (e) WS; and (f) BA.}
 \label{fig: Budget_influence}
\end{figure*}

\begin{figure*}[h]
\centering
    \includegraphics[width=0.8\linewidth]{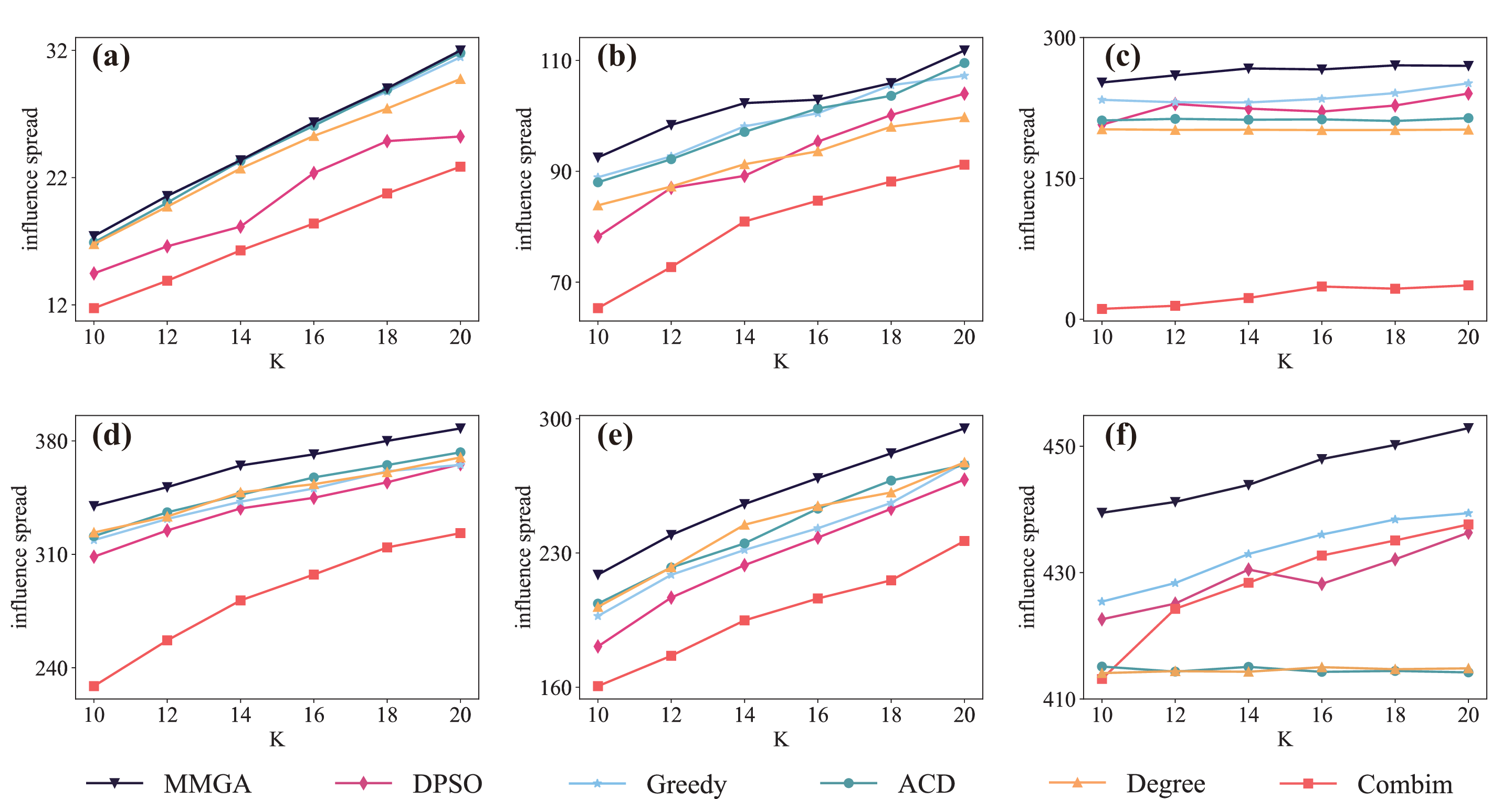}
 % \captionsetup{font=small} % 设置字体大小为小号
 \caption{The influence spread of different algorithms under various seed set sizes, in which we set $B=600$. The results are given for networks: (a) Transport; (b) CKM; (c) Drosophila; (d) ER; (e) WS; and (f) BA.}
 \label{fig: K_influence}
\end{figure*}

\noindent
The performance of the algorithms under different budgets and seed sizes is shown in Figure~\ref{fig: Budget_influence} and Figure~\ref{fig: K_influence}, respectively. In Figure~\ref{fig: Budget_influence}, we set the seed size as $K=20$, and we set the budget as $B=600$ in Figure~\ref{fig: K_influence}. Figure~\ref{fig: Budget_influence} shows that MMGA could always find the optimal seed set to maximize the spread of influence with different budget values $B$. It performs better than the baseline methods, especially for small values of $B$.  DPSO performs second best on a small budget in most networks. However, MMGA outperforms DPSO in all networks by almost $10\%$ when $B \leq 300$. Greedy only performs relatively well in Transport network, but shows unstable performance in other networks. For degree-based benchmarks, i.e., ACD and Degree, MMGA achieves an average improvement of $12.8\%$ and $15\%$, respectively. The selection of seeds of Combim starts from the community with the smallest size, and once the number of nodes reaches $K$, it can no longer choose additional nodes. In other words, the budget is not fully utilized, resulting in the algorithm's poor performance.
Overall speaking, the spread of influence in different networks is relatively stable for MMGA, whereas most of the baselines show an increasing trend in most networks with the increase in budget. This implies that the baseline methods are more sensitive to budget changes compared to MMGA. Similar to Figure~\ref{fig: Budget_influence}, MMGA also performs best when we change the seed set size constraint $K$ in different networks, as shown in Figure~\ref{fig: K_influence}. With an increase in $K$, the spread of influence increases for most algorithms. The performance of the baselines is consistent with those in Figure~\ref{fig: Budget_influence}.

To further explore the robustness of MMGA, we increase the probability of propagation of the MIC model to larger values, such as $p=0.3$ and $0.5$, and keep the budget and seed size as $B=600$ and $K=20$, respectively. The spread of influence of the seed set obtained by different algorithms is given in Table~\ref{tab:differ_p}, where the best performance is shown in bold, and the second best is indicated with an underline. Generally speaking, when the propagation probability is high, the strong spreading coupling of multilayer networks narrows the performance differences between different algorithms, which is particularly evident in dense synthetic networks, such as WS and BA networks. Furthermore, the table shows that MMGA outperforms the other benchmarks under MIC model regardless of how $p$ is set. Meanwhile, Combim achieves the second-best result, with ACD following closely behind. Both the Combim and ACD algorithms fundamentally adopt the concept of reducing influence overlap, their advantages becoming more apparent as the propagation probability increases. 

\begin{table*}[h]%调节图片位置，h：浮动；t：顶部；b:底部；p：当前位置
	\centering
	\caption{The influence spread of different algorithms under various propagation probability and we set $K=20$ and $B=600$. The best performance is shown in bold and the second best is shown with an underline in each multilayer network.\\}
	\label{tab:differ_p} 
    \resizebox{0.9\linewidth}{!}{
	\begin{tabular}{|c|c|c|c|c|c|c|c|c|c|c|c|c|}%表格中的数据居中，c的个数为表格的列数
		\hline\hline\noalign{\smallskip}	
		& \multicolumn{6}{|c|}{p=0.3} & \multicolumn{6}{|c|}{p=0.5} \\
		\noalign{\smallskip}\hline\noalign{\smallskip}
         & Transport & CKM & Drosophila & ER & WS & BA & Transport & CKM & Drosophila & ER & WS & BA\\
        \noalign{\smallskip}\hline\noalign{\smallskip}
        MMGA & \textbf{68.91} & \textbf{228.57} & \textbf{1617.63} & \textbf{965.75} & \textbf{981.12} & \textbf{947.35} & \textbf{125.37} & \textbf{239.36} & \textbf{2390.26} & \textbf{996.01} & \textbf{999.56} & \textbf{995.45} \\ 
        \noalign{\smallskip}\hline\noalign{\smallskip}
        DPSO & 54.09 & 221.30 & \underline{1604.02} & 957.27 & 980.73 & 945.01 & 114.36 & 235.84 & 2372.94 & 992.03 & 999.51 & 995.24 \\
        \noalign{\smallskip}\hline\noalign{\smallskip}
        Greedy& 61.67 & 221.51 & 1594.19 & 957.62 & 980.85 & 946.02 & 113.92 & 235.82 & 2363.28 & 991.80 & \underline{999.56} & 995.30\\
        \noalign{\smallskip}\hline\noalign{\smallskip}
        ACD & \underline{67.39} & 221.06 & 1587.89 & 957.13 & 980.20 & 944.46 & \underline{119.02} & 236.01 & 2359.59 & 991.79 & \underline{999.56} & 995.26\\
        \noalign{\smallskip}\hline\noalign{\smallskip}
        Degree & 58.67 & 221.08 & 1585.15 & 957.61 & 980.81 & 943.85 & 104.91 & 235.94 & 2361.98 & 991.95 & 999.47 & 995.16\\
        \noalign{\smallskip}\hline\noalign{\smallskip}
        Combim & 37.72 & \underline{222.35} & 1596.63 & \underline{959.59} & \underline{980.94} & \underline{946.32} & 79.64 & \underline{236.38} & \underline{2374.88} & \underline{992.79} & 999.46 & \underline{995.45}\\
		\noalign{\smallskip}\hline
	\end{tabular}
    }
\end{table*}

In the initialization of the population in MMGA, we use a combination of degree, influence-cost ratio (ICR), and random strategy to generate the populations. We investigate the influence of this combination on the final performance of MMGA, in which we show the AUC values of different initialization methods, i.e., Mixed (combination of degree, influence-cost ratio (ICR), and random strategy), Degree, ICR, and Random, when keeping $K$ unchanged. The AUC value is the area under the influence spread curve when the value of the budget $B$ changes. As shown in the table, the method that combines Degree, SRC, and Random initialization consistently achieves the best results, highlighted in bold, in different settings of seed set sizes. In most cases, the random strategy achieves the second-best performance. Actually, using the combination of the three different strategies cannot only achieve better performance in spread influence, but also could help to accelerate convergence speed of MMGA.

\begin{table}[h]%调节图片位置，h：浮动；t：顶部；b:底部；p：当前位置
	\centering
	\caption{The AUC value under different initializations. The best performance is shown in bold and the second best is shown with an underline in each capacity constrain $K$.\\}
	\label{tab:initialization} 
    \resizebox{0.7\linewidth}{!}{
	\begin{tabular}{cccccccc}%表格中的数据居中，c的个数为表格的列数
		\hline\hline\noalign{\smallskip}	
		Dataset& Initialization & K=10 & K=12  & K=14 &K=16 & K=18 &K=20\\
		\noalign{\smallskip}\hline\noalign{\smallskip}
    \multirow{4}{*}{Transport}     & Mixed &\textbf{0.2582} &\textbf{0.2569} &\textbf{0.2548} &\textbf{0.2532} &\textbf{0.2543} &\textbf{0.2548} \\
	   & Degree &\underline{0.2470} &0.2473 &0.2469 &0.2489 &0.2483 &0.2475\\
                & ICR  &0.2464 &0.2466 &0.2486 &0.2484 &0.2478 &0.2490\\
                & Random &0.2484 &\underline{0.2492} &\underline{0.2496} &\underline{0.2495} &\underline{0.2496} &\underline{0.2487}\\
         \noalign{\smallskip}\hline\noalign{\smallskip}
    \multirow{4}{*}{CKM}    & Mixed &\textbf{0.2509} &\textbf{0.2519} &\textbf{0.2524} &\textbf{0.2519} &\textbf{0.2527} &\textbf{0.2512}\\
         & Degree &\underline{0.2506} &0.2495 &0.2490 &0.2512 &0.2447 &0.2498\\
            & ICR &0.2487 &0.2483 &0.2484 &\underline{0.2515} &\underline{0.2513} &0.2490\\
            & Random &0.2499 &\underline{0.2503} &\underline{0.2502} &0.2454 &\underline{0.2513} &\underline{0.2500}\\
        \noalign{\smallskip}\hline\noalign{\smallskip}
        
    \multirow{4}{*}{Drosophila}    & Mixed &\textbf{0.2536} &\textbf{0.2531} &\textbf{0.2517} &\textbf{0.2503} &\textbf{0.2513} &\textbf{0.2505}\\
     & Degree &0.2475 &0.2481 &0.2495 &0.2501 &0.2482 &0.2498\\
               & ICR &0.2488 &\underline{0.2500} &0.2489 &0.2494 &0.2499 &0.2496\\
               & Random &\underline{0.2501} &0.2489 &\underline{0.2499} &\underline{0.2502} &\underline{0.2506} &\underline{0.2501}\\
        \noalign{\smallskip}\hline\noalign{\smallskip}
        
    \multirow{4}{*}{ER}   & Mixed &\textbf{0.2535} &\textbf{0.2545} &\textbf{0.2533} &\textbf{0.2527} &\textbf{0.2519} &\textbf{0.2516}\\
     & Degree &0.2487 &\underline{0.2492} &0.2487 &0.2491 &0.2491 &\underline{0.2493}\\
               & ICR &0.2478 &0.2478 &0.2483 &0.2487 &0.2491 &0.2500\\
               & Random &\underline{0.2501} &0.2485 &\underline{0.2497} &\underline{0.2496} &\underline{0.2500} &0.2491\\
        \noalign{\smallskip}\hline\noalign{\smallskip}

  \multirow{4}{*}{WS}  & Mixed &\textbf{0.2588} &\textbf{0.2584} &\textbf{0.2578} &\textbf{0.2570} &\textbf{0.2557} &\textbf{0.2552}\\
    & Degree &0.2452 &0.2450 &0.2458 &0.2479 &0.2485 &0.2469\\
               & ICR &0.2466 &0.2478 &0.2469 &0.2460 &0.2468 &0.2478\\
               & Random &\underline{0.2495} &\underline{0.2488} &\underline{0.2495} &\underline{0.2491} &\underline{0.2490} &\underline{0.2501}\\
          \noalign{\smallskip}\hline\noalign{\smallskip}

    \multirow{4}{*}{BA}   & Mixed &\textbf{0.2503} &\textbf{0.2501} &\textbf{0.2505} &\textbf{0.2506} &\textbf{0.2501} &\textbf{0.2506}\\
     & Degree &\underline{0.2500} &\underline{0.2500} &0.2496 &0.2499 &0.2500 &0.2497\\
               & ICR &0.2497 &0.2499 &0.2498 &0.2494 &0.2498 &0.2498\\
               & Random &\underline{0.2500} &\underline{0.2500} &\underline{0.2501} &\underline{0.2501} &\underline{0.2501} &\underline{0.2499}\\

		\noalign{\smallskip}\hline
	\end{tabular}
    }
\end{table}

\section{Conclusion}
\noindent
Given that the current budgeted influence maximization problem only considers the limitation of budget, which may result in unreasonable solutions that are not applicable to real-world scenarios, we propose a new influence maximization problem in multilayer networks that considers both the budget and seed size, i.e., BCIM problem. The challenges for solving this problem are mainly as follows: (\romannumeral 1) How to design an algorithm that meets the two constraints in BCIM? (\romannumeral2) How to find a seed set that is influential in different layers? To this end, we propose a multilayer multi-population genetic algorithm (MMGA), which uses repair and muti-population parallel modules to solve the budget and seed size constraints, respectively. In addition, the design of the initialization module and the fitness function in MMGA that considers nodes' properties and influence in multilayer networks can not only help to find the optimal influential seed set, but also navigate the algorithm to converge quickly. Meanwhile, a spreading model, i.e., MIC, which considers a piece of information spreading through a multilayer network and enhances inter-layer spread interactions, is proposed to quantify the spread influence of the nodes selected by different methods. To validate the effectiveness of our algorithm, we perform experiments on three artificial and three empirical multilayer networks from different domains. The results show that the proposed algorithm is superior to state-of-the-art benchmarks in terms of influence spread and robustness. Furthermore, through a detailed parameter analysis, we found that the crossover module and prior knowledge of initialization play crucial roles in the MMGA framework.

This paper presents the first attempt at solving the BCIM problem in multilayer networks and opens up a series of future research directions. One is to design efficient algorithms when the costs of nodes are assigned in a more complex manner rather than solely based on node degree. Secondly, the multilayer independent cascade model used in this work is a simplification of the real-world scenario, which is similar to the independent cascade model on an aggregated weighted network of the corresponding multilayer network~\cite{41}. In future work, one could consider a more realistic case where the spreading processes are characterized by different models to solve the BCIM problem. Besides, the fitness score of each node in the MMGA framework is obtained by Monte Carlo simulations, which have a relatively high computational complexity, especially on large-scale networks. Therefore, how to estimate the influence spread of each node in a multilayer network could also be a promising direction for exploration. Finally, we plan to apply the MMGA algorithm to real-world digital marketing scenarios to gain valuable insight into promoter selection strategies. This application has the potential to improve resource allocation efficiency and increase overall returns on investment.

\section{Acknowledgement}
This work was supported by the Natural Science Foundation of Zhejiang Province (Grant No. LQ22F030008), the Natural Science Foundation of China (Grant No. 61873080) and the Scientific Research Foundation for Scholars of HZNU (2021QDL030).

\renewcommand\refname{\large

\end{document}